\documentstyle[11pt,twoside,jltp,graphicx]{article}

\title{On developed superfluid turbulence}

\author{G.E. Volovik\address{Low Temperature Laboratory,
Helsinki University of Technology,
P.O.Box 2200, FIN-02015 HUT, Finland\\
L.D. Landau Institute for
Theoretical Physics,  Kosygin Str. 2, 117940 Moscow, Russia}}

\runninghead{G.E. Volovik}{On developed superfluid turbulence}

\begin{document}

\begin{abstract}
  Superfluid turbulence is governed by two
dimensionless parameters. One of them is the intrinsic parameter
$q$ which characterizes the relative value of the friction force acting on
a vortex with respect to the non-dissipative forces. The inverse
parameter $q^{-1}$ plays the
same role as the  Reynolds number ${\rm Re}=UR/\nu$ in classical
hydrodynamics. It marks the transition between the "laminar" and turbulent
regimes of vortex dynamics. The developed turbulence, described by
a Kolmogorov cascade, occurs when ${\rm Re}\gg 1$ in classical
hydrodynamics. In superfluids, the developed turbulence occurs at $q\ll
1$. Another parameter of superfluid turbulence is the superfluid
Reynolds number
${\rm Re}_s=UR/\kappa$, which contains the circulation quantum $\kappa$
characterizing quantized vorticity in superfluids. The two parameters $q$
and ${\rm Re}_s$ control the crossover or transition between two classes
of superfluid turbulence: (i) the classical regime,
 where the Kolmogorov cascade (probably modified by the non-canonical
dissipation due to mutual friction) is effective, vortices are locally
polarized, and the quantization of vorticity is not important; and (ii)
the Vinen  quantum turbulence where the properties are determined by the
quantization of vorticity. The phase diagram of these dynamical vortex
states is suggested.

\end{abstract}

\maketitle

\section{Introduction}

Recent experiments in $^3$He-B \cite{Finne} demonstrated that
the fate of few vortices injected into a rapidly moving
superfluid depends on dimensionless
intrinsic temperature dependent parameter $q$, rather
than on the flow velocity. At $q$ of order unity a rather sharp
transition is observed
between the laminar evolution of the injected vortices and the emerging
turbulent many-vortex state of the whole superfluid.
This added a new twist to the general theory of turbulence in superfluids
developed by Vinen \cite{Vinen,VinenNiemela} and others. Attempts to
modify the  theory in order to incorporate the new phenomenon, have been
made in Refs. \cite{Kopnin}  and \cite{Volovik2003}.
Here we follow the latter work \cite{Volovik2003}
where we utilized the coarse-grained hydrodynamic equation for the
dynamics of the superfluid with distributed vortices.  In this equation
the parameter $q$ characterizes  the friction force between the superfluid
and normal components of the liquid, which is mediated by quantized
vortices. According to this equation, turbulence develops only if the
friction is relatively small compared to the inertial term,
i.e. when $q$ is below unity.
We also argued that the developed turbulence must occur at $q\ll 1$,
and suggested that in this region, where the inertial term is dominating,
there are at least two possible states
of turbulence. One of them corresponds to the state discussed by Vinen,
where the microscopic nature of quantized vortices is essential. While in
the other state, turbulence does not depend on the circulation
quantum, and thus the information on the underlying microscopic
physics of quantum liquid is lost. Turbulence in this state becomes
similar to turbulence in classical liquids, and it is also  described
by the Richardson cascade at least at the initial stage of the
development. However, as distinct from the classical liquids, the final
state of turbulence is determined not by viscosity  but by the mutual
friction parameter $q$.

These differences from the classical turbulence arise because the
hydrodynamics of superfluid liquid exhibits new features with respect to
conventional classical hydrodynamics, which become important when
turbulence in superfluids is considered:

(i) The superfluid liquid consists of mutually penetrating components
--  the viscous normal
component and one or several frictionless superfluid components.
That is why different types of turbulent motion are
possible depending on whether the normal and superfluid components move
together or separately. Here we are interested in the most simple case
when the dynamics of the normal component can be neglected. This occurs,
for example, in the superfluid phases of $^3$He where the normal component
is so viscous that it is practically clamped to the container walls. The
role of the normal component in this case is to provide the preferred
heat-bath reference frame, where the normal component and thus the heat
bath are at rest. The dissipation takes place when vortices move with
respect to this reference frame. The turbulence in the superfluid
component with the normal component at rest is  here referred to as
superfluid turbulence. We also assume that there is only a single
superfluid component. For
$^3$He-B this means that we  ignore the spin degrees of freedom, assuming
that all three superfluid spin components  move together.

(ii) The important feature of  superfluid turbulence is that the
vorticity of the superfluid component is quantized in terms of the
elementary circulation quantum $\kappa$ (in $^3$He-B $\kappa=\pi\hbar/m$,
where $m$ is the mass of $^3$He atom). So, superfluid turbulence is
the chaotic motion of well determined and well separated vortex filaments
\cite{VinenNiemela}. Using this as starting point we can simulate the main
ingredients of classical turbulence -- the chaotic dynamics of the
vortex degrees of freedom of the liquid. However, to make the analogy
useful one must choose the regime where the microscopic nature of the
superfluid -- the circulation quantum containing the Planck constant
$\hbar$ and  the mass of the atom $m$ -- is not involved.

(iii) The most important distinction from the classical hydrodynamics is
that the dissipation in vortex motion is not due to the viscosity term
$\nu\nabla^2{\bf v}$ in the Navier-Stokes
equation. The superfluid component does not exhibit viscosity, instead
the dissipation occurs due to the friction force acting on the superfluid
vortex when it moves with respect to the heat bath (the normal
component). The force acting on a single vortex is proportional to
the velocity of the vortex in the heat-bath frame. In the coarse-grained
hydrodynamics of the superfluid with the distributed vorticity, this
leads to the force between the superfluid and normal components, which is
proportional to their relative velocity ${\bf v}_{\rm s}- {\bf v}_{\rm
n}$, and to the vortex density
$|\nabla\times{\bf v}_{\rm s}|$. As a result the dissipative term in the
hydrodynamic equation is non-linear, and its structure reminds that of
the non-linear inertial term. The relative magnitude of the two
non-linear terms, the friction and inertial ones, is given by the
dimensionless parameter $q$. Thus the quantity $1/q$ plays the role of the
Reynolds number. This is an internal parameter of the liquid, as distinct
form the Reynolds number in conventional liquids where it depends on the
flow velocity and dimension of the system. When $q\ll 1$, the inertial
term is dominating, and this corresponds to a big Reynolds number in
classical liquids. In  this regime one can expect, that the turbulent
state obeys the Richardson energy cascade, governed by the inertial term.
We discuss here whether the Kolmogorov scaling law survives the
non-canonical dissipation in superfluids, i.e. whether or not the
scaling is modified due to the non-linear dissipative term.

\section{Coarse-grained hydrodynamic equation}

The coarse-grained hydrodynamic equation is
obtained from the Euler equation for the superfluid velocity  ${\bf
v}\equiv {\bf v}_{\rm s}$  after averaging over the vortex lines
(see Refs. \cite{Hall} and review \cite{Sonin}). Instead of the
Navier-Stokes equation with
the $\nu\nabla^2{\bf v}$ term one has
\begin{eqnarray}
\frac{\partial {\bf v} }{ \partial t}+ \nabla\mu=  {\bf
v} \times
  \vec{\bf\omega} ~-
\label{SuperfluidHydrodynamics1}
\\
  -\alpha'({\bf
v} -{\bf v}_{\rm
n})\times
  \vec{\bf\omega}+  \alpha~\hat{\bf\omega}
\times(\vec{\bf\omega}
\times({\bf
v} -{\bf v}_{\rm
n}) ) ~.
\label{SuperfluidHydrodynamics2}
\end{eqnarray}
Here ${\bf v}_{\rm n}$  is the velocity of the normal component;
$\vec{\bf\omega}=\nabla\times {\bf v}$ is the superfluid vorticity;
$\hat\omega=\vec{\bf\omega}/\omega$; and the dimensionless parameters
$\alpha'$ and $\alpha$ come from the reactive and dissipative forces
acting on a vortex when it moves with respect to the normal component.
These parameters are very similar to the Hall and conventional
resistivities,
$\rho_{xy}$ and $\rho_{xx}$, in the Hall effect. For vortices in fermionic
systems (superfluid $^3$He and superconductors) they were calculated by
Kopnin
\cite{KopninBook}, and measured in $^3$He-B over a broad temperature
range by Bevan et. al.
\cite{Bevan} (see also \cite{VolovikBook}, where these parameters
are discussed in terms of the chiral anomaly). 
We omitted the higher order
terms, including the term discussed by Hall \cite{Hall} which contains
the bending energy of vortex lines (the condition for that is discussed
in Section \ref{Crossover} where another dimensionless parameter will be
introduced -- the superfluid Reynolds number
${\rm Re}_s=UR/\kappa$ which  contains the circulation quantum $\kappa$).

The terms in
expression (\ref{SuperfluidHydrodynamics1}) are invariant with respect to
the transformation ${\bf v}\rightarrow {\bf v}({\bf r}-{\bf u}t) +{\bf
u}$ as in classical hydrodynamics. However, the terms in
expression (\ref{SuperfluidHydrodynamics2}) are not invariant under this
transformation, since there is the
preferred reference frame in which the normal component is at rest.
They are invariant under the full Galilean transformation
when the normal component is also involved: ${\bf v}\rightarrow {\bf
v}({\bf r}-{\bf u}t) +{\bf u}$ and
${\bf v}_{\rm n}\rightarrow {\bf v}_{\rm n} +{\bf u}$.

The experiments in Ref.\cite{Finne} were made in a rotating cryostat,
where the normal component exhibits solid body rotation,
  ${\bf v}_{\rm n}={\bf \Omega}\times {\bf r }$. However, here we study
the local properties of turbulence and avoid this complication,
assuming that
  ${\bf v}_{\rm n}$ is uniform. Actually this means that we consider the case of
strong turbulence where the vortex density essentially exceeds the equilibrium
vortex density in the rotating container: $\omega\gg \Omega$.
We shall work in the frame comoving with the normal component, where
${\bf v}_{\rm n}=0$, but we must remember that this frame is unique.
In this frame the equation for superfluid hydrodynamics is simplified:
\begin{equation}
\frac{\partial {\bf v} }{ \partial t}+ \nabla\mu= (1-\alpha'){\bf
v} \times
  \vec{\bf\omega}+  \alpha~\hat{\bf\omega} \times(\vec{\bf\omega}
\times{\bf v} ) ~.
\label{SuperfluidHydrodynamics}
\end{equation}
After rescaling the time,
$\tilde t=(1-\alpha')t$, one obtains an equation
\begin{equation}
\frac{\partial {\bf v} }{ \partial \tilde t}+ \nabla \tilde\mu=
  {\bf v} \times
  \vec{\bf\omega}+  q~\hat{\bf\omega} \times(\vec{\bf\omega}
\times{\bf v} ) ~,
\label{SuperfluidHydrodynamics3}
\end{equation}
which depends on a single
parameter $q$:
\begin{equation}
q=\frac{\alpha}{1-\alpha'} ~.
\label{q}
\end{equation}
Now in Eq.(\ref{SuperfluidHydrodynamics3}) the first three terms together
are the same as the inertial terms in classical hydrodynamics. They
satisfy the modified Galilean invariance:
\begin{equation}
  {\bf v}( \tilde t,{\bf r}) \rightarrow {\bf v}( \tilde t,{\bf r}-{\bf
u}\tilde t)+{\bf u} ~.
\label{GalileanInvariance}
\end{equation}
In fact the transformation above
changes the chemical potential, but this does not influence the vortex
degrees of freedom which are important for the  phenomenon of
turbulence.
  In contrast, the
dissipative last term with the factor
$q$ in Eq.(\ref{SuperfluidHydrodynamics3}) is not invariant under this
transformation.  This is in
contrast to the conventional liquid where the whole Navier-Stokes
equation, which contains the kinematic viscosity
\begin{equation}
\frac{\partial {\bf v}}{ \partial t}+ \nabla\mu=  {\bf v}\times
  \vec{\bf\omega}+  \nu\nabla^2 {\bf v}
~,
\label{NormalHydrodynamics}
\end{equation}
is Galilean invariant, and where there is
no preferred reference frame.

This difference between the dissipative last
terms in Eqs. (\ref{NormalHydrodynamics}) and
(\ref{SuperfluidHydrodynamics3}) is very important:

(1) The role of the Reynolds number, which characterizes the ratio of
the inertial and dissipative terms in the hydrodynamic equations,  is
played by the intrinsic parameter
$1/q$ in superfluid turbulence. This parameter does not depend on the
characteristic velocity $U$ and size
$R$ of the large-scale flow as distinct from the conventional
Reynolds number ${\rm
Re}=RU/\nu$ in classical viscous hydrodynamics. That is why the
turbulent regime occurs only at
$1/q>1$ even if vortices are injected to the superfluid which moves with
large velocity $U$. This rather unexpected result was obtained in
experiments with superfluid $^3$He-B \cite{Finne}.

(2) In conventional
turbulence the large-scale velocity $U$ is always understood as the
largest characteristic velocity difference in the inhomogeneous
flow of the classical
liquid \cite{McComb}.
In the two-fluid system the velocity
$U$ is  the large-scale velocity of the
superfluid component with respect to the normal component, and this
velocity (the so-called counterflow velocity) can be completely
homogeneous (however, for the intermediate turbulent state obtained
in the $^3$He-B
experiments in rotating vessel, the large-scale velocity field is
inhomogeneous and we use
$U\sim\Omega R$, with $\Omega$ being the rotation velocity and $R$ -- the
radius of the container).

(3) As distinct from the classical hydrodynamics,  the energy
dissipation is produced by the non-linear mutual friction (mf) term -- the
last term in Eq.(\ref{SuperfluidHydrodynamics3}):
\begin{equation}
  \epsilon_{\rm mf} =-\dot E=-\langle{\bf v}\cdot \frac{\partial {\bf v}
}{
\partial
\tilde t}\rangle=  -q \langle {\bf v}\cdot(\hat{\bf\omega}
\times(\vec{\bf\omega}
\times{\bf v} ))\rangle \sim  q \omega
v^2 ~.
\label{EnergyDissipation}
\end{equation}
This reflects the fact that the mutual friction depends on the velocity of
the superfluid component with  respect to the normal one, i.e. on the
absolute value of the velocity
$v$ in the frame of the container. Thus for a given value of vorticity
$\omega$ the largest dissipation  occurs at the large-scale
velocity
$U$. On the other hand, according to the Kolmogorov scaling, the maximum
vorticity
$\omega$ is concentrated at the smallest scale. This gives the nontrivial
dependence of the dissipation on scales: dissipation is not concentrated
at the smallest scale only.

(4)  The onset of the superfluid turbulence was studied by Kopnin in
Ref. \cite{Kopnin}. His Vinen-type model demonstrated that
the initial avalanche-like multiplication of vortices leading to
turbulence occurs when $q$ drops below unity, which is in agreement
with experiment
\cite{Finne}.
  The existence of two regimes in the initial development of vorticity is
also supported by earlier simulations
by Schwarz who noted that when $\alpha$ (or $q$) is reduced a
crossover from a regime of isolated phase slips to a phase-slips cascade
and then to the fully developed vortex turbulence occurs
\cite{Schwarz}.

(5) One can expect that the well developed turbulence occurs when
$q\ll 1$, and here we shall discuss this extreme limit. In
$^3$He-B the condition $q\ll 1$ is realized at temperatures well
below
$0.6T_c$
\cite{Finne}. However, we do not consider very low temperatures where
instead of the mutual friction the other mechanisms of dissipation take
place such as excitation of Kelvin waves
\cite{KelvinWaveCascade} and vortex reconnection
\cite{TsubotaArakiNemirovskii}. The latter leads to formation of cusps
and kinks on the vortex filaments whose fast dynamics creates a burst
of different types of excitations in quantum liquids: phonons, rotons,
Kelvin waves and fermionic quasiparticles. A similar  burst of
gravitational waves from cusps and kinks on cosmic strings was
theoretically investigated by the cosmological community (see e.g.
\cite{DamourVilenkin}), and the obtained results are very important
for the superfluid turbulence at a very low temperature.

(6) We expect that even at  $q\ll 1$, two
different states of turbulence are possible, with the crossover (or
transition) between them being determined by $q$ and by another
dimensionless parameter
${\rm Re}_s=UR/\kappa$, where $\kappa$ is the circulation around the
quantum vortex. The coarse-grained hydrodynamic equation
(\ref{SuperfluidHydrodynamics3}) is in fact valid only in the limit
${\rm Re}_s\gg 1$, since the latter means that the characteristic
circulation of the velocity $\Gamma=UR$ of the large-scale
flow substantially exceeds the circulation quantum $\kappa$, and thus
there are many vortices in the turbulent flow. When ${\rm Re}_s$
is reduced the quantum nature of vortices becomes more pronounced. At
some (still big) value of ${\rm Re}_s$ we proceed from the type of
superfluid turbulence which resembles the classical turbulence  and is
probably described by the (modified) Kolmogorov cascade, to the quantum
regime which is probably described by the Vinen equations for the average
vortex dynamics \cite{Vinen}.

\section{Kolmogorov cascade}
\label{KolmogorovCascade}

Let us first consider the possibility of a Kolmogorov-Richardson
cascade in superfluid turbulence.
Let us start with the analysis based on the dimensionalities and on  the
idea of the energy cascade. In the next Section we shall use a more
detailed analysis based on model dynamical equations for the energy
spectrum and on dispersion of dissipation over the scales.

In classical
turbulence, a large Reynolds number
${\rm Re}=UR/\nu\gg 1$ leads to well separated length scales or wave
numbers. As a result the Kolmogorov-Richardson cascade takes place in
which the energy flows from small wave numbers $k_{min}\sim 1/R$ (large
rings of size $R$ of the container) to high wave number $k_0=1/r_0$ where
the dissipation occurs.  In the same manner in our case of superfluid
turbulence the necessary condition for a Kolmogorov cascade is a large
ratio of the inertial and dissipative terms in
Eq.(\ref{SuperfluidHydrodynamics3}), i.e.
$1/q\gg 1$.

In the  Kolmogorov-Richardson cascade,
at arbitrary length scale  $r$ the energy transfer rate to
the smaller scale, say $r/2$, is $v_r^2/ t_r$, where $v_r^2$
is the kinetic energy at this scale, and $t_r=r/v_r$ is the
characteristic time  (turn-over time). If there are no appreciable
losses of energy in the intermediate scales, the energy transfer from one
scale to the next must be the same for all scales. As a result one has
\begin{equation}
  \epsilon= \frac{v_r^2}{ t_r}=\frac{v_r^3}{ r}={\rm
constant}=\frac{U^3}{ R}~.
\label{Micriscopic1}
\end{equation}
 From this equation it follows that
\begin{equation}
v_r=  \epsilon^{1/3}~ r^{1/3}~.
\label{Micriscopic2}
\end{equation}
This certainly occurs in classical liquids, and there is some indication
that this might take place in superfluids too
\cite{TsubotaKasamatsuAraki}.  However, in superfluids
the dissipation mechanism is different, which can change the overall
picture: even if initially the energy flows to the smaller scale according
to the Richardsonin, this does not mean that in the final steady state the
Kolmogorov scaling will be established.

In classical liquids, the energy dissipation due to viscosity is
$\epsilon_{\rm visc}\sim \nu (\nabla_i{\bf v})^2$. It is concentrated at
the smallest scale $r_0$, which is found from
the energy balance $\epsilon=\epsilon_{\rm visc}\sim\nu  v_{0}^2/
r_0^2=\nu\epsilon^{2/3}r_0^{-4/3}$. This gives 
\begin{equation}
    r_0={1/k_0} \sim   R {\rm Re}^{-3/4}~.
\label{MinimalScaleClass}
\end{equation}

In superfluid turbulence, the
situation is more complicated. According to Eq.(\ref{EnergyDissipation}),
in superfluids  the dissipation is not a linear function of the velocity.
In the non-linear theory, as follows from the Fourier analysis,
contributions from different scales are not independent. Since the
vorticity
$\omega_r=v_r/r$ is concentrated  at the smallest scale, while the
velocity -- at the largest scales, one obtains that the overall
dissipation due to mutual friction is $\epsilon_{\rm mf} \sim  q
\omega_{0} U^2$. Let us make a natural assumption that the Kolmogorov
cascade stops when the overall dissipation becomes comparable to the
energy transfer from scale to scale $\epsilon$ in Eq.(\ref{Micriscopic1})
\begin{equation}
\epsilon =\epsilon_{\rm mf}\sim q \omega_{0} U^2 \sim qU^2 \frac{v_{0}}{
r_0}=
  qU^2 \epsilon^{1/3}~ r_0^{-2/3} \sim    \frac{U^3}{ R}~.
\label{EnergyDissipation2}
\end{equation}
Then, from the above equation one obtains the smallest length scale in the
superfluid turbulence:
\begin{equation}
    r_0={1/k_0} \sim  q^{3/2}R~,
\label{MinimalScale}
\end{equation}
which depends on the internal parameter $q$ instead of the Reynolds
number ${\rm Re}$ in Eq.(\ref{MinimalScaleClass}) for classical liquids.
The associated characteristic velocity $v_{0}$ and vorticity $\omega_0$ at
this scale are
\begin{equation}
  v_{0}\sim q^{1/2}U~,~\omega_0=k_0v_0 \sim
{U\over qR}~.
\label{VelVortMinimalScale}
\end{equation}
This consideration is valid when $r_0\ll R$ and $v_{0}\ll U$, which
means that $1/q\gg 1$ is the condition for the Kolmogorov cascade. In
classical liquids the corresponding condition for well developed
turbulence is
${\rm Re}\gg 1$. In both cases these conditions ensure that the kinetic
terms in the hydrodynamic equations are much larger than the
dissipative terms in a large enough inertial range. In the same manner as
in classical liquids the condition for the stability of the turbulent flow
is ${\rm Re}>1$, one may suggest that the condition for the stability of
the discussed turbulent flow is $1/q>1$. This is
supported by observations in $^3$He-B where it was demonstrated that
at high velocity $U$ but at
$q>1$, turbulence is not developed even after vortices were
introduced into the flow
\cite{Finne}.

 In the Kolmogorov cascade
both in classical liquids and superfluids the kinetic
energy is concentrated at large scales comparable to the container size:
\begin{equation}
  E=\int_{r_0}^R\frac{dr}{ r} v_r^2=\int_{r_0}^R\frac{dr}{ r} (\epsilon
r)^{2/3}= (\epsilon R)^{2/3} =U^2~.
\label{KineticEnergyR}
\end{equation}

The dispersion of the turbulent energy in the momentum space is also the
same as in classical liquid
\begin{eqnarray}
  E= \int_{r_0}^R\frac{dr}{ r} v_r^2=\int_{k_0}^{1/R}{dk\over k}
v_k^2=\int_{k_0}^{1/R} \frac{dk}{ k}  \frac{\epsilon^{2/3}}{ k^{2/3}}
=\int_{k_0}^{1/R} dk E(k),
\nonumber
\\
   E(k)=v_k^2/k=\epsilon^{2/3} k^{-5/3}.
\label{KineticEnergyK2}
\end{eqnarray}

However, the dispersion of dissipation caused by the mutual friction
is different from that in viscous classical liquids, where the main
dissipation occurs at the smallest scale. This can change the whole
pattern of the turbulent steady state. 

\section{Dispersion of dissipation}

 In superfluids the direct
transfer of the kinetic energy to the normal component due to the mutual
friction occurs at all scales. Let us consider the dispersion $
\epsilon_{r}^{\rm mf}$ of this dissipation as a function of the length
scale
$r$. In the nonlinear term $q\omega v^2$ the  given scale $r$ contributes
through the velocity field ${\bf v}$ and the vorticity field $|\omega|$.
The most important contributions are from the fluctuating vorticity
$\omega_r$ and the fluctuating velocity $v_r$:
\begin{equation}
 \epsilon_{r}^{\rm mf}=
\epsilon_{r}^{(1)}+\epsilon_{r}^{(2)}=q\langle {v^2\over\omega}\rangle 
\omega_r^2
 +q\langle \omega \rangle v_r^2  ~.
\label{TwoDissipationTerms}
\end{equation}

The term $\epsilon_{r}^{(1)}$ comes from the dissipation experienced by 
the fluctuating vorticity $\omega_r$
at scale $r$. Assuming the Kolmogorov scaling one obtains the following
estimation of this term
\begin{equation}
 \epsilon_{r}^{(1)}=q\langle {v^2\over\omega}\rangle 
\omega_r^2=q{\langle  v^2 \rangle \over\langle\omega\rangle} 
\omega_r^2=
{qU^2\over \omega_0} \omega_r^2=\epsilon (r_0/r)^{4/3} ~,
\label{Diss1}
\end{equation}
 Here as before $\epsilon=U^3/R$ is the energy flux from
scale to scale in the inertial range. We take into account that the
vorticity moves with respect to the normal component with the
characteristic velocity $U$ and thus experiences the corresponding mutual
friction. The Eq.(\ref{Diss1}) corresponds to the effective turbulent
viscosity:
 $\epsilon_{r}^{(1)}=\nu_{eff}\omega_r^2$ with $\nu_{eff}=qU^2/\omega_0$.
As in the classical liquids this dissipation is peaked at the smallest
scale where the vorticity is concentrated.

The second term in Eq.(\ref{TwoDissipationTerms}) is the contribution of
the velocity fluctuations to the friction experienced by the
average vorticity (see Eq.(16) in Ref. \cite{VinenNew}). Under the same
assumption of the Kolmogorov scaling, the second term has the following
magnitude:
\begin{equation}
 \epsilon_{r}^{(2)}=q\omega_0 v_r^2=\epsilon (r/R)^{2/3} ~.
\label{Diss2}
\end{equation}
Though the average vorticity is concentrated at the smallest scale,
$\langle \omega \rangle=\omega_0$, the term
$\epsilon_{r}^{(2)}$ as a function of
$r$ is peaked at largest scale $r=R$ where the velocity and kinetic
energy are concentrated. This term which removes energy at large scales
was also considered in classical liquids (see e.g. Eq.(5) in Ref.
\cite{Toschi} with $\alpha=q\omega_0$).

The other possible terms in the dissipation are smaller than these two if
$q\ll 1$. For example, one can add the term
$qv_r^2\omega_r$. In the Kolmogorov cascade it does not depend on
the scale $r$, and is always much smaller than the sum of the two
main terms:
$q\omega_r v_r^2=q\epsilon \ll q^{2/3}\epsilon={\rm
min}_r(\epsilon_{r}^{(1)}+\epsilon_{r}^{(2)})$.
Thus at $q\ll 1$ we have two peaks where the dissipation is concentrated.
These two peaks at two extreme scales, $r_0$ and $R$,
have equal magnitude if the Kolmogorov scaling is obeyed:
\begin{equation}
 \epsilon_{r=r_0}^{(1)}\sim \epsilon_{r=R}^{(2)}\sim\epsilon=U^3/R ~.
\label{Diss12}
\end{equation}

Let us now assume that dissipation which occurs at a given scale $r$
is on the order of the losses of the kinetic energy at the same scale,
i.e.
$\partial_tE_r\sim -\epsilon_r$. Since
the dissipation is nonlinear, this is not the fact; nevetheless,
this is a rather natural assumption. If one accepts this at least for the
two scales, $R$ and $r_0$, then using the double-peak structure of the
dispersion of the dissipation one comes to the following scenario of the
cascade.  The part of the initial kinetic energy of the superflow at
large scale
$r=R$ is directly transferred to the normal component of the liquid due
to the mutual friction in Eq.(\ref{Diss2}).  The remaining comparable part of the
flow energy experiences the Kolomogorov-Richardson cascade to the smaller
scales until the next dissipation peak due to Eq.(\ref{Diss1}) is reached
at
$r=r_0$.

\section{Possible modifications of the Kolmogorov cascade due to
dispersion of dissipation}

If our assumption is correct, the above scenario with two peaks of
dissipation does not change the Kolmogorov scaling discussed
in Sec.
\ref{KolmogorovCascade}, though some part of the kinetic energy is
dissipated already at large scale. However, in principle, the dispersion
of the dissipation can modify the energy cascade and the scaling law. Let
us consider this using the following model of dissipation, which probably
is not very realistic but allows us to find the modified scaling law
within the model. Let us suppose that the two peaks overlap so that
instead of the double-peak structure, the dissipation due to the direct
transfer of the energy to the normal component is equally distributed
over the scales:
\begin{equation} \epsilon_r^{\rm mf} =\epsilon^{\rm mf} \sim q
\omega_{0} U^2~,
 \label{EnergyDissipation3}
 \end{equation}
and consider, whether the picture of the Kolmogorov-Richardson cascade
survives or
not under this new input. And if yes, how this modifies the
Kolmogorov spectrum.
Due to the direct losses of energies at each scale the energy
transfer from scale to scale $\epsilon_r^{\rm cascade}$ decreases with
decreasing scale, so that the Eq.(\ref{Micriscopic1}) does not hold any
more:
\begin{equation}
  \epsilon_r^{\rm cascade}= \frac{v_r^2}{ t_r}=\frac{v_r^3}{ r} \neq
{U^3\over R}\equiv \epsilon~
\label{Micriscopic2}
\end{equation}
Instead we must write an equation for the $r$-dependent
$\epsilon_r^{\rm cascade}$ in the steady state, which takes into account
that the energy flux from the scale $r$ to the neighbouring scale $r/2$
splits into the direct energy transfer from the scale $r/2$ to the normal
component and the energy flux to the next scale: $\epsilon_r^{\rm
cascade}=\epsilon_{r/2}^{\rm cascade}+\epsilon_{\rm mf}$. Thus the
difference between the energy fluxes at neighbouring scales is balanced
by  the direct energy losses at each step in
Eq.(\ref{EnergyDissipation3}) and one obtains the following equation for
$\epsilon_r^{\rm cascade}$:
\begin{equation}
r\partial_r \epsilon_r^{\rm cascade}= \epsilon_{\rm mf}~,~~{\rm or}~~~ r\partial_r \left({v_r^3\over r}\right)=
\epsilon_{\rm mf}~.
\label{EqForEpsilon}
\end{equation}
In terms of wave numbers $k=1/r$ and taking
into account the time
dependence,  the energy balance in Eq.(\ref{EqForEpsilon}) can be written
as
\begin{equation}
\partial_t (v_k^2)+k\partial_k (kv_k^3)=
- \epsilon^{\rm mf}~,~{\rm or}~~~
\partial_t E(k)+\partial_k (k^{5/2}E^{3/2})=
- {\epsilon_{\rm mf}\over k}~.
\label{EqForEpsilonK}
\end{equation}
where as before the energy $E(k)=v_k^2/k$.

We can also use the more sophisticated equations of Leith type
describing the diffusion of energy in the $k$-space, which
  have been considered for turbulent cascades in conventional liquids
  \cite{Leith}. For example, in a recent publication
\cite{ConnaughtonNazarenko}  the following diffusion equation was used:
\begin{equation}
\partial_t E = {1\over 8}\partial_k \left(
k^{11/2}E^{1/2}\partial_k(E/k^2)\right)   +f-\nu k^2E ~.
\label{DiffusionModelTurbulence}
\end{equation}
Here $f$ is an external source of energy.
In superfluids, instead of the last (viscous) term,
there is a sink of energy caused by the
direct transfer of energy to the normal component by mutual
friction. In our model with constant dispersion of dissipation one has:
\begin{equation} \partial_t E = {1\over 8}\partial_k \left(
k^{11/2}E^{1/2}\partial_k(E/k^2)\right)  
 +f -{1\over k}\epsilon_{\rm mf}~. 
\label{DiffusionModelSuperfluidTurbulence} \end{equation} 
For this model
the steady-state solutions of equations (\ref{EqForEpsilonK}) and
(\ref{DiffusionModelSuperfluidTurbulence}) are the same (assuming that
the source term $f(k)=0$ at $k>1/R$):
\begin{equation}
E(k)= C \left(\epsilon_{\rm mf}  \right)^{2/3} k^{-5/3}
\ln^{2/3} {k_0\over k}~.
\label{DiffusionModelSolution}
\end{equation}
They differ only by the prefactor $C$ which is  $C=(24/11)^{2/3}$ for
Eq.(\ref{DiffusionModelSuperfluidTurbulence}), and $C=1$
for Eq.(\ref{EqForEpsilonK}).

Eq.(\ref{DiffusionModelSolution}) demonstrates the main
difference in the energy spectrum
for the conventional and modified Kolmogorov cascades.  In both cases the
energy spectrum can be represented as
 \begin{equation} E(k)= C
\epsilon^{2/3} k^{-5/3}F(k/k_0)~.  \label{GeneralForm}
\end{equation}
But in
conventional turbulence one has $F(k\ll k_0)=1$, while in the modified
turbulence the function $F(k/k_0)$ logarithmically diverges at $k_0\gg k$.

 From Eq.(\ref{DiffusionModelSolution}) it follows that the velocity
\begin{equation}
v_k=(kE)^{1/2} \sim (\epsilon_{\rm mf})^{1/3}
k^{-1/3}\ln^{1/3} {k_0\over k}
\label{NonKolmVelocity}
\end{equation}
  monotonically decreases with $k$ and approaches zero at
$k\rightarrow k_0$. On the other hand the vorticity
\begin{equation}
\omega_k=kv_k\sim (\epsilon_{\rm mf})^{1/3}k^{2/3}\ln^{1/3} {k_0\over k}
\label{NonKolmVorticity}
\end{equation}
  first increases, then reaches its maximum
\begin{equation}
\omega_{max}\sim (\epsilon_{\rm mf})^{1/3}k_0^{2/3}
\label{MaximumVorticity}
\end{equation}
  at
$k_{max}=k_0/\sqrt{e}$ and finally almost abrubtly goes to zero within
the same scale $k_0$.

The kinetic energy $E(k)$ decreases first as $k^{-5/3}$. Then, when the
scale
$k_0$ is reached, it vanishes at $k=k_0$ as
$(k_0-k)^{2/3}$. When $k$ approaches $k_0$, the non-linear
inertial term in the hydrodynamic equation decreases and approaches the
non-linear friction term. The latter becomes dominating and the cascade stops.

Let us now find how the value of $k_0$ in Eq.(\ref{MinimalScale})
is modified by the logarithmic function.
The direct transfer of the energy to the heat bath $\epsilon_{\rm mf}$
is determined by the maximal vorticity:
\begin{equation}
  \epsilon_{\rm mf} \sim  q
\omega_{max} U^2~.
\label{DiffusionModelEnergyLosses2}
\end{equation}
The total energy losses $\epsilon$ is given by
\begin{equation}
\epsilon=  \int_{1/R}^{k_0}  {dk\over k}\epsilon_{\rm mf}=
\epsilon_{\rm mf}\ln (k_0R)~,
\label{DiffusionModelEnergyLosses}
\end{equation}
As a result the energy spectrum becomes
\begin{equation}
  E(k)=C\epsilon^{2/3} k^{-5/3} {\ln^{2/3} {k_0\over k}\over \ln^{2/3}
k_0R}.
\label{DiffusionModelSolution3}
\end{equation}

 From equations
(\ref{MaximumVorticity}), (\ref{DiffusionModelEnergyLosses}) and
(\ref{DiffusionModelEnergyLosses2}) it follows that
$\omega_{max}\sim q^{1/2}k_0U\sim \epsilon/(qU^2\ln(k_0R))$.
Then using the equation for energy at large scale
$R$, $U^2=E(k=1/R)/R$, one obtains from Eq.(\ref{DiffusionModelSolution3})
that the total energy losses to the heat bath is $\epsilon=U^3/R$.
Finally one comes to the logarithmic modification of
equation (\ref{MinimalScale}) for the scale $k_0$
\begin{equation}
   k_0={1\over r_0} ~~~,
~~ r_0=R q^{3/2}\ln (k_0R)\approx  R q^{3/2} \ln(1/q)~.
\label{DiffusionModelSolutionk}
\end{equation}
The characteristic velocity $v_0$ at this scale $r_0$
  and vorticity $\omega_{max}$ are:
\begin{equation}
\omega_{max}\sim {U\over Rq\ln(1/q)}~, v_0={\omega_{max}\over
k_0}\sim q^{1/2}U~.
\label{DiffusionModelSolutionv0}
\end{equation}

Another model of the dispersion of the dissipation has been recently
considered by Vinen in Ref. \cite{VinenNew}. Vinen took into account 
a single term in dissipation,  Eq.(\ref{Diss2}), which is peaked at
large scale. He solved the
Eq.(\ref{DiffusionModelSuperfluidTurbulence}) with this type of
dissipation and obtained the solution with a stronger modification of the
Kolmogorov cascade.

Here we discussed the steady state solutions of diffusion equation, but in
principle, the
time-dependent diffusion equation (\ref{DiffusionModelTurbulence}) for
$E(k,t)$ can be used for the analysis of the formation and decay of the
turbulent state with different dispersion of the dissipation.

\section{Crossover to Vinen quantum turbulence}
\label{Crossover}

Assuming that the superfluid turbulence is described by the non-modified
Kolmogorov cascade, let us discuss the two possible regimes which occur
at different ranges of $q$. At a very small $q$ the microscopic nature of
vortices with quantization of circulation becomes important. The
condition of the validity of the coarse-grained hydrodynamic description
used above is that the circulation relevant in the turbulent state can be
considered as continuous.  This means that the circulation at the
smallest scale $r_0$ must be still larger than the circulation quantum,
$v_{0}r_0> \kappa$:  \begin{equation} v_{0} r_0 =U Rq^2 =q^2 \kappa {\rm
Re}_s  > \kappa ~~,~~{\rm Re}_s=\frac{UR}{\kappa}~.
\label{PhaseBoundary1} \end{equation} Thus the constraint for the validity of
the Kolmogorov cascade is \begin{equation} {\rm Re}_s > \frac{1}{  q^2 }\gg 1
~ .  \label{PhaseBoundary2} \end{equation} The same condition can be derived
from the requirement that the characteristic scale $r_0$ must be much larger
that the intervortex distance $l$. The latter is obtained from the vortex
density in the Kolmogorov state \begin{equation} n_{\rm
  K}=\frac{\omega_0}{\kappa}  \sim   \frac {U }{  \kappa R q} = \frac{1}{
R^2} \frac{{\rm Re}_s}{ q}~~.  \label{VortexDensity} \end{equation} The
  condition $l\ll r_0$ leads again to the equation $v_{0}r_0> \kappa$ and
thus to the criterion (\ref{PhaseBoundary2}).  Actually the same condition
allows us to neglect the term in the coarse grained equation one can neglect
the term which takes into account the energy of the bending of vortex lines
\cite{Hall}.

In connection to Eq.(\ref{VortexDensity}) let us mention the experiments on
$^3$He-B in the rotating vessel
\cite{Finne}. There the turbulent state appears as the intermediate
state between the
initial vortex-free state and the final state where the superfluid
experiences in
average the solid-body rotation ${\bf v}={\bf \Omega}\times {\bf r}$, where
${\bf \Omega}$ is the angular velocity of the rotating vessel. In the final
equilibrium state the vortex density is $n_{\rm equilibrium}=2\Omega/\kappa$. From
Eq.(\ref{VortexDensity}) and taking into account that $\Omega=U/R$
one obtains that in
the intermediate turbulent state the vortex density must exceed the equilibrium
density:
\begin{equation}
  n_{\rm K}\sim {1\over  q}n_{\rm equilibrium}~.
\label{VortexDensityVsEquil}
\end{equation}
The excess of vortices in the intermediate state has been observed in
the experiment
\cite{Finne} and also in numerical simulations of this experiment
\cite{Tsubota}.

\begin{figure}[!!!t]
   \centerline{\includegraphics[width=0.8\linewidth]{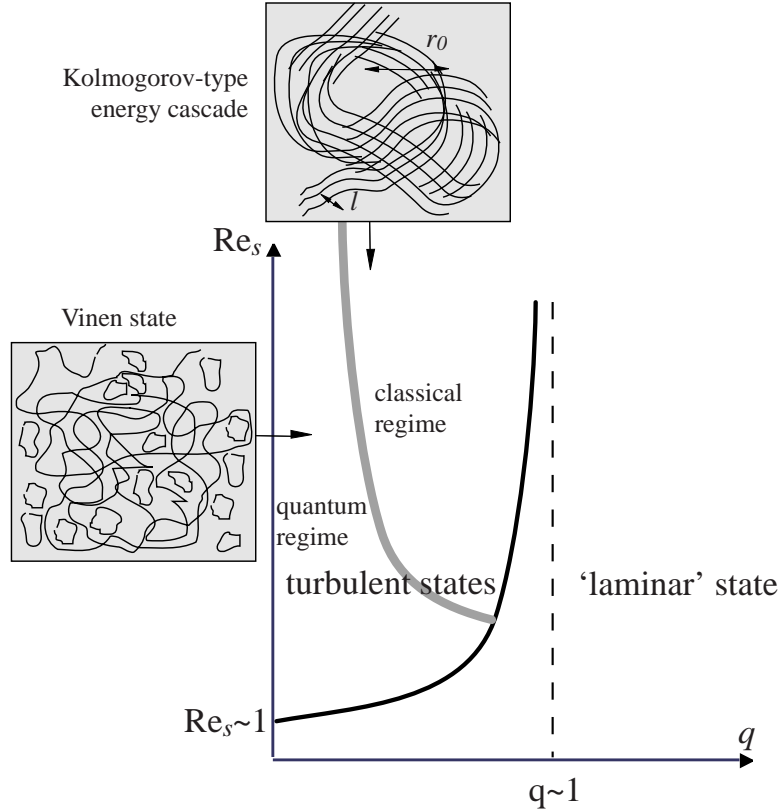}}
   \caption{Possible phase diagram of dynamical vortex states in the $({\rm
Re}_s,q)$ plane. At large flow velocity ${\rm Re}_s\gg 1$, the boundary
between turbulent and `laminar' vortex flow approaches the vertical
axis
$q=q_0\sim 1$ as suggested by experiment \protect\cite{Finne}. The thick
line separates two regimes of superfluid turbulence occurring at small
$q$:  (i) the developed turbulence of the classical type, which is
characterized by the Kolmogorov-type cascade possibly modified by the
mutual friction; and (ii) the quantum turbulence of the Vinen type at even
smaller $q$. }
   \label{PhaseDiagram}
\end{figure}

The criterion (\ref{PhaseBoundary2}) contains the `superfluid
Reynolds number' ${\rm
Re}_s$, which is determined by the microscopic characteristic of the
vortex state
  -- the circulation quantum $\kappa$. This
superfluid Reynolds number is responsible for the crossover
or transition from the classical superfluid turbulence, where the
quantized vortices are locally aligned (polarized) forming thick vortex tubes
(Fig. 1~{\it top}),
and thus the quantization is not important, to the quantum
turbulence of isolated quantized vortex lines (Fig. 1~{\it left})
whose description was developed by Vinen.

We can now consider the approach to the crossover from the quantum
regime -- the Vinen state which probably occurs when ${\rm Re}_s
q^2<1$.
According to Vinen \cite{Vinen} the turbulence in
the quantum regime is characterized by a
single scale.  The characteristic velocity is the counterflow velocity
$U$, while the characteristic length scale $l$ is the distance between the
vortices or the size of the characteristic vortex loops. It is
determined  by the circulation quantum and the counterflow velocity,
$l=\lambda
\kappa /U$, where
$\lambda$ is the dimensionless intrinsic parameter, which probably
contains
$\alpha'$ and
$\alpha$. The vortex density in the Vinen state is
\begin{equation}
  n_{\rm V}=l^{-2}\sim \lambda^2     \frac{U^2}{\kappa^2}=
\frac{\lambda^2}{ R^2}  {\rm Re}_s^2 ~,
\label{VortexDensityVinen}
\end{equation}
It differs from the vortex density in the Kolmogorov state in
Eq.(\ref{VortexDensity}) which depends not only on the counterflow
velocity $U$, but also on the container size $R$.
If the crossover between the classical and quantum regimes of the
turbulent states occurs at ${\rm Re}_s q^2=1$, the two equations
(\ref{VortexDensityVinen}) and (\ref{VortexDensity}) must match each other
in the crossover region. But this occurs only if
$\lambda^2\sim q$. If $\lambda^2\neq q$ there is a mismatch, and one may
expect that either the two states are separated by the first-order phase
transition, or there is an intermediate region where the superfluid
turbulence is described by two different microscopic scales such as
$r_0$ and
$l$.

In any case the crossover between the classical and quantum
regimes occurs at the border of applicability of the coarse-grained
equation (\ref{SuperfluidHydrodynamics}), i.e. we cannot use this
equation to describe the Vinen state. For this state, in which the vortex
lines are not locally aligned, the Vinen equations must be used. Note
that there is a principle difference  between these two turbulent states,
which comes from the different dependence of the vortex density on $U$
and $R$ in Eq.(\ref{VortexDensityVinen}) and Eq.(\ref{VortexDensity}).
In particular, in the Vinen state, the mutual friction force between
the normal and superfluid components (the last term in
Eq.(\ref{SuperfluidHydrodynamics3})) obeys the  cubic law suggested by
Gorter and Mellink \cite{GorterMellink1949}:
$f\sim q
\kappa n_{\rm V} U \sim q\lambda^2 U^3/\kappa$. In the Kolmogorov state,
the friction force is obeying the quadratic law, $f\sim qU\omega_0
\sim U^2/R$.

Based on the above consideration one may suggest the phase
diagram of different regimes of collective vortex dynamics in Fig. 1.
It is possible that the boundary between the `laminar' and
`turbulent' regions reflects the process of developing of the turbulent
vortex cluster: in the  `laminar' region turbulence cannot be started
by a few injected vortices, while in the turbulent region the injected
vortices lead to the vortex avalance and the turbulent state
\cite{Kopnin}. Actually this is what was observed in the experiment
\cite{Finne}. However, the more fundamental role of this boundary, as the
line of the phase transition between the vortex states, is not excluded:
it is possible that in the laminar region the steady-state turbulence is
simply unstable and decays. The instability of the turbulent vortex
cluster in the lower part of the phase diagram was observed by Schwarz in
his numerical simulations,  but whether the boundary between the turbulent
and laminar vortex flow approached the vertical axis $q=q_0\sim 1$ at
large velocity as suggested by experiment \cite{Finne}, was not clear
from the simulations
\cite{VinenPrivate}.

\section{Discussion}

Superfluid turbulence (as well as turbulence in classical
3-dimensional liquids)
is a collective many-vortex phenomenon. Here we discussed the simple
case  when the normal component is at rest,
or its motion is fixed (which actually occurs in $^3$He-B). Also we considered
the dynamics of a single superfluid component, i.e.
we ignored the other possible superfluid components in the
multi-component superfluids,
such as the spin degrees of freedom in $^3$He-B: spin currents and
spin vortices.
But even in this simple case there can be several
different types of collective dynamical vortex states.
Each of these vortex states can be
characterized by its own correlation functions. For example,
as the characteristic which distinguishes between different vortex
states one can use
the behavior of the loop function
\begin{equation}
g(C)=\left<e^{i(2\pi/\kappa)\oint_C {\bf v}\cdot d{\bf r}}
\right>~.
\label{LoopFunction}
\end{equation}
In the limit when the length $L$ of the loop $C$ is much larger than
the intervortex distance $l$ one may expect the general behavior $g(L)\sim
\exp(-(L/l)^{\gamma})$ where the exponent
$\gamma$ is different for different vortex states.
The asymptotic behavior of the loop function has been used for the
description of the
equilibrium phase transition in condensed matter and quantum field
theory (see e.g.
Ref. \cite{Toulouse1979,Polyakov1987}). It can be used also for the
description of the
non-equilibrium phase transition \cite{VolovikBook} where vortices
also appear in the
intermediate state according to the Kibble-Zurek mechanism \cite{Kibble,Zurek}.

In principle, the parameters $\alpha$ and $\alpha'$
may depend on the type of the dynamical state, since they are obtained by
averaging of the forces acting on individual vortices. The
renormalization of these parameters $\alpha(L)$ and $\alpha'(L)$, when the
length scale $L$ is increasing,
  may also play an important role in the identification of the turbulent
states, as in the case of the renormalization-group flow of similar
parameters $\rho_{xx}$ and $\rho_{xy}$ in the quantum Hall effect (see e.g.
\cite{Khmelnitskii}).

One can expect phase transitions between different states of collective
vortex dynamics. One of such transitions, which appeared to be rather
sharp, has been observed between the `laminar' and `turbulent' dynamics of
vortices in superfluid $^3$He-B \cite{Finne}. It was found that this
transition was regulated by the intrinsic velocity-independent
dimensionless parameter
$q=\alpha/(1-\alpha')$, rather than by flow velocity.
However, it is not excluded that
both dimensionless parameters $\alpha$ and $\alpha'$ are important. Moreover,
it is also possible that only the initial stage of the formation of the
turbulent state is governed by these parameters \cite{Kopnin}.

Another
transition (or maybe a crossover) is suggested here between the
quantum and classical regimes of developed superfluid
turbulence. We argue that there is a range of parameters in the classical
region of the $({\rm Re}_s,q)$ plane where turbulence of vortex lines is
described by the Kolmogorov-Richardson cascade, or by some modification
of it caused by the dispersion of the dissipation.
 In this regime,  turbulence is similar to that in classical
liquids with modified dissipation. Thus the superfluid serves as a
physically motivated example of the liquid with the non-canonical
dissipation, which requires the general analysis of different models
of dissipation and forcing (see e.g. \cite{Toschi}). 

The turbulence in classical liquids is thought to be
characterized by the dynamics of the vortex tubes,  whose radii are of
order of the dissipative Kolmogorov scale (see e.g. Ref.
\cite{Laval} and references therein). In superfluid turbulence, the radii
of such tubes  are determined by the parameter $q$ acording to
Eq.(\ref{DiffusionModelSolutionk}):
$r_0\sim Rq^{3/2}$. In superfluids, the classical description is valid
when these vortex tubes are fat enough, i.e. they contain many quantized
vortices. The circulation around the tubes essentially exceeds the
circulation quantum $\kappa$, so that we can ignore the quantum nature of
superfluid vortices. In this region we can study analytically, 
numerically and experimentally all the phenomena which take place in
classical liquids, including intermittency.

The crossover to the quantum regime occurs when the circulation around
the relevant vortex tubes becomes comparable to the circulation quantum
$\kappa$.  According to Vinen, in the quantum regime, turbulence is
represented by a single length scale -- the intervortex distance $l$ --
and the single velocity scale
$U$, which are related through the circulation quantum: $l\sim \kappa/U$.
It was suggested by Skrbek \cite{Skrbek} that such a crossover between the
quantum Vinen state and the classical Kolmogorov state of superfluid
turbulence was probably observed in several experiments with the
counterflowing superfluid $^4$He. Skrbek identified these states with
turbulent states I and II according to the Tough's classification scheme
\cite{Tough}.

  \section*{ACKNOWLEDGMENTS}

I thank V.B. Eltsov, D. Kivotides, N.B. Kopnin, M. Krusius, V.V.
Lebedev,  L. Skrbek, F. Toschi,
M. Tsubota and W.F. Vinen for
fruitful discussions and criticism.  This work was supported by
ESF COSLAB Programme and by the Russian Foundation for
Fundamental Research.


\begin{thebibliography}{15}

\bibitem{Finne} A.P. Finne, T. Araki, R. Blaauwgeers,
V.B. Eltsov, N.B. Kopnin, M. Krusius, L. Skrbek, M. Tsubota,
and G.E. Volovik,  An intrinsic velocity-independent criterion for
superfluid turbulence, Nature {\bf 424}, 1022--1025 (2003).

\bibitem{Vinen} W.F. Vinen, Proc. R. Soc.. London, ser A {\bf 242}, 493
(1957).

\bibitem{VinenNiemela} W.F. Vinen, and J.J. Niemela,   J. Low Temp.
Phys.   {\bf 128}, 167 (2002).

\bibitem{VinenNew} W.F. Vinen, Superfluid turbulence in the presence of a stationary
normal fluid, to be published in Phys. Rev. B

\bibitem{Kopnin} N.B. Kopnin, Vortex instability and the onset of
superfluid turbulence, Phys. Rev. Lett. to be published; cond-mat/0309708.

\bibitem{Volovik2003} G.E. Volovik, `Classical and quantum regimes of
the superfluid turbulence',     Pisma ZhETF {\bf 78}, 1021--1025  (2003)
\lbrack JETP Letters {\bf 78},   533--537 (2003)\rbrack.

\bibitem{Hall} H.E. Hall, in: {Liquid Helium, International School of Physics "Enrico Fermi" Course
XXI}, G. Carreri, ed. , Academic Press, New York (1963); Phil. Mag. Supplement {\bf 9}, issue 33, 89 (1960).

\bibitem{Sonin} E.B. Sonin, Rev. Mod. Phys. {\bf 59}, 87 (1987).

\bibitem{KopninBook} N.B. Kopnin,    {\it Theory of
Nonequilibrium Superconductivity}, Clarendon Press, Oxford (2001).

\bibitem{Bevan} T.D.C. Bevan {\it et al.},
Nature,  {\bf 386},  689  (1997); J. Low Temp. Phys, {\bf 109}, 423 (1997).

\bibitem{VolovikBook} G.E. Volovik, {\it The Universe in a
Helium Droplet}, Clarendon Press,  Oxford (2003).


\bibitem{McComb}  W.D. McComb, {\it The Physics of Fluid Turbulence},
Clarendon Press, Oxford (1990).


\bibitem{Schwarz} K.W. Schwarz,  Fluid dynamics of a quantized vortex
filament in a hole, Physica {\bf B~197}, 324 (1994); Numerical experiments
on single quantized vortices, preprint.

\bibitem{KelvinWaveCascade} W. F. Vinen, Makoto Tsubota, and Akira
Mitani, Kelvin-wave cascade on a vortex in superfluid $^4$He at a very low
temperature, Phys. Rev. Lett. {\bf 91},  135301 (2003).

\bibitem{TsubotaArakiNemirovskii} Makoto Tsubota, Tsunehiko Araki and
Sergey K. Nemirovskii, Dynamics of vortex tangle without mutual  friction
in superfluid $^4$He, Phys. Rev.  {\bf B~62}  11751-11762 (2000).

\bibitem{DamourVilenkin} Thibault Damour and Alexander Vilenkin,
Gravitational wave bursts from cusps and kinks on cosmic strings,
      Phys.Rev. {\bf D~64},  064008 (2001).

\bibitem{Toschi} L. Biferale, M. Cencini, A. S. Lanotte, 
M. Sbragaglia and F. Toschi, Anomalous scaling and universality in
hydrodynamic systems with power-law forcing,
             New J. Phys. {\bf 6}, 37 (2004) .

\bibitem{Leith} C. Leith, Phys. Fluids {\bf 10}, 1409 (1967);
Phys. Fluids {\bf 11}, 1612 (1968).


\bibitem{ConnaughtonNazarenko} Colm Connaughton and Sergey
Nazarenko,       Warm Cascades and Anomalous Scaling in a Diffusion Model
of Turbulence, Phys. Rev. Lett. {\bf 92}, 044501 (2004).

\bibitem{TsubotaKasamatsuAraki} Makoto Tsubota, Kenichi Kasamatsu,
Tsunehiko Araki, Dynamics of quantized vortices in
superfluid helium and rotating Bose-Einstein condensates,
cond-mat/0309364.

\bibitem{Tsubota} Makoto Tsubota, private communications.

\bibitem{VinenPrivate}  W.F. Vinen, private communications.

\bibitem{GorterMellink1949}  C.J. Gorter and  J.H. Mellink, On
the irreversible processes in liquid helium II, Physica {\bf 15},
285--304  (1949).


\bibitem{Toulouse1979} G. Toulouse,  Gauge concepts in condensed
matter physics,  in: {\it Recent
Developments In Gauge Theories}, 331--362 (1979).

\bibitem{Polyakov1987}  A.M. Polyakov  {\it Gauge Fields and
Strings}, Contemporary Concepts in Physics {\bf 3}, Harwood Academic,
Chur (1987).


\bibitem{Kibble}  T.W.B. Kibble, Topology of cosmic domains
and strings,  J. Phys.  {\bf A~9},   1387--1398    (1976).

\bibitem{Zurek} W.H. Zurek, Cosmological experiments in
superfluid helium?, Nature {\bf 317}, 505  (1985).


\bibitem{Khmelnitskii} D.E. Khmelnitskii, Quantization of Hall
conductivity, Pis'ma ZhETF {\bf 38}, 454--458 (1983).

\bibitem{Laval} J.-P. Laval, B. Dubrulle, S. Nazarenko,
Non-locality and Intermittency in 3D Turbulence,
Phys. of Fluids {\bf 13},  1995--2012 (2001).

\bibitem{Skrbek} L. Skrbek, Flow phase diagram for the helium superfluids,
cond-mat/0402301.

\bibitem{Tough} J.T. Tough, Superfluid turbulence, in: {\it Prog. in
Low Temp. Phys.},
vol. VIII, North Holland, Amsterdam (1982).



\end{thebibliography}
\end{document}